# THE IMPACT OF COVID-19 ON THE UK FRESH FOOD SUPPLY CHAIN

**Rebecca Mitchell[1], Roger Maull[2], Simon Pearson[3], Steve Brewer[3] and Martin Collison[4]**

1. University of Exeter Business School, Rennes Drive, Exeter, EX4 4PU

2. Initiative for the Digital Economy, University of Exeter, Paris Gardens, London, SE1 8ND, UK

3. Lincoln Institute of Agri-Food Technology, The University of Lincoln, Lincoln, LN6 7TS, UK

4. Collison and Associates Ltd, Shepherdsgate Rd, Tilney All Saints, King's Lynn PE34 4RW

**ABSTRACT:**

The resilience of the food supply chain is a matter of critical importance, both for national security and broader societal well-being. COVID-19 has presented a test to the current system, as well as means by which to explore whether the UK's food supply chain will be resilient to future disruptions. In the face of a growing need to ensure that food supply is more environmentally sustainable and socially just, COVID-19 also represents an opportunity to consider the ability of the system to respond innovatively, and its capacity for change. The purpose of this case-based study is to explore the response and resilience of the UK fruit and vegetable food supply chain to COVID-19, and to assess this empirical evidence in the context of a resilience framework based on the adaptive cycle. To achieve this we reviewed secondary data associated with changes to retail demand, conducted interviews with 23 organisations associated with supply to this market, and conducted four video workshops with 80 organisations representing 50% of the UK fresh produce community. The results highlight that, despite significant disruption, the retail dominated fresh food supply chain has demonstrated a high degree of resilience. The role of industry bodies, levy boards and innovation communities has been important, but the focus of the response has been on maintaining the current just-in-time operating model. In the context of the adaptive cycle, the system has shown signs of being stuck in a rigidity trap, as yet unable to exploit more radical innovations that may also assist in addressing other drivers for change. This has highlighted the significant role that innovation and R&D communities will need to play in enabling the supply chain to imagine and implement alternative future states post-COVID.

**MAIN TEXT**

**1. INTRODUCTION / MOTIVATION**

The COVID-19 global pandemic has affected virtually every section of society including the food supply chain. This is critically important as we all need to eat, and challenging, as the food sector comprises a rich and diverse mix of large, small and micro sized businesses all interacting in a vibrant, if muddy, globally distributed ecosystem. Media articles with headlines such as "*rationing in the UK is inevitable*" (Telegraph, 2020), "*COVID-19 pandemic risks worst global food crisis in decades*" (NewScientist, 2020), and "*Hunger crisis hits lockdown Britain…*" (Daily Mail, 2020) have reported on a wide range of challenges including food poverty, labour shortages and the implementation of "*draconian measures*" (Daily Mail 2020). In many cases however the articles are presenting a complex mix of issues and potential outcomes; at this stage there is relatively little evidence of systematic review.

In many developed economies, in normal times there is a highly efficient just-in-time distribution system, largely controlled by a small number of powerful retailers supported by an army of niche operators, often operating on very tight margins (Reardon & Timmer, 2012). For the consumer this means that food is relatively cheap; UK households spend just c. 10% of their expenditure on food (Defra, 2020a). However, the efficiency and complexity of the supply chain also mean that it is susceptible to external shocks. The impact of COVID-19 on the food supply chain therefore presents a challenge and an opportunity for food businesses. The challenge is to survive the disruption initiated by the virus, and to mitigate its impacts; the opportunity is to learn and apply approaches that build increased resilience, and decreased fragility, to future disruption.

The EPSRC Internet of Food Things Network Plus has been working alongside the food sector since 2018 to investigate how emerging technologies and innovation can enhance the digitalisation of the supply chain. This has highlighted potential for digital innovation to play a significant role in a number of areas including reducing food waste, increasing nutritional value, increasing productivity and reducing environmental impacts. Prior to COVID-19, challenges associated with supply chain resilience to significant external disruption had not been raised explicitly, though some aspects of resilience may be addressed by innovation interventions aimed primarily at addressing other challenges. R&D expenditure by businesses in the UK Food & Beverage and Agriculture & Fishing sectors is low relative to many other sectors (Office for National Statistics, 2019), and uptake of new technologies has tended to focus on incremental innovation and refining existing business practices, rather than on radical ideas and investments that fundamentally change the operating practices within the sector.

Therefore, there is huge potential value in understanding the extent to which the sector has been affected by COVID-19, to evaluate the resilience that the sector has exhibited and to consider the appetite for more radical innovations that will increase resilience for the future. At a top level, our key research questions were:

- What was the change in food demand resulting from COVID-19?
- What was the impact on food suppliers?
- What types of innovation are emerging?

The food supply chain can be broken down in a number of ways, notably by tier (e.g. producer, manufacturer, retailer), by product (e.g. fruit & veg, dairy, meat), by location of origin (e.g. UK grown or imported), and by final destination (e.g. retail, food service sector). The operating practices, stressors and drivers for each of these categories can vary considerably. For the purposes of this initial study we have focused our exploration of changes to food demand on the retail sector as opposed to the service sector as changes to demand within the service sector are likely to be predominantly due to the rapid shut-down of much of this sector by enforceable Government restrictions (Cabinet Office, 2020). We have chosen to focus our exploration of impact on food suppliers primarily on UK producer and processing of fruit, vegetables and potatoes. The reasons for this are multiple:

- The UK is already reliant on imports of fruit and vegetables – with only 17% self-sufficiency for fresh fruit, and 52% self-sufficiency for fresh vegetables (Defra, 2020b). If UK production in these areas were to collapse due to a lack of resilience, further reducing self-sufficiency, there may be negative implications for food security in the event that imports are also affected by factors such as trade agreements (Brexit) and border restrictions (COVID). This is particularly relevant in light of continued campaigns to improve the health of the nation through increased consumption of fruit and veg (e.g. NHS Eatwell, 2017; Willet et al, 2019)
- Much of the produce from this sub-sector has limited shelf-life, therefore traditional mechanisms to increased supply-chain resilience (such as increasing stocks as part of inventory management) are not applicable. More innovative approaches are likely to be necessary.
- The Internet of Food Things Network Plus has had extensive previous engagement with actors within this sub-sector, in part representing the nature of agriculture and food processing sectors within Greater Lincolnshire.

Combining these factors, we have focused on dominant sectors by economic size – namely the supermarket coordinated supply chains, although we have also examined some of the alternative providers and channels including farm shops and online deliveries. Therefore, our final research questions are:

1. What was the change in retail demand for fruit and vegetables resulting from COVID-19?
2. What was the impact on UK suppliers of fruit and vegetables?
3. What types of innovation are emerging within supermarket coordinated supply chains of fruit and vegetables?

This paper therefore seeks to present empirical evidence on the impact within the scope we have defined, and an exploration of the impact and emerging innovations drawn from our interpretation of this data and interviews with actors across the supply chain. We consider the current situation in the context of previous empirical and conceptual studies of supply chain resilience, and reflect upon the role of innovation and R&D practitioners in improving future resilience.

## 2. LITERATURE REVIEW & RESEARCH FRAMING

### 2.1. Supply chain management and the food sector

Inspired by events such as the fuel protests in the UK in 2000, the outbreak of Foot and Mouth Disease in the UK in 2001, and the 9/11 terrorist attacks in the USA in 2001, the early 2000's saw a number of studies into supply chain vulnerability and risk management (Juttner et al, 2003; Tang, 2006). In the UK, a number of practical guides were published, providing Business Continuity Management focused tools, and supporting risk managers hitherto focused on management of risk internal to a firm (e.g. Christopher & Peck, 2003; Christopher & Rutherford, 2004; Sheffi & Rice, 2005). Shortly after the identification of Food as a national infrastructure sector in the 2004 Civil Contingencies Act, the Department of Food, Environment and Rural Affairs commissioned a study of Food Supply Chain Resilience in the UK (Peck, 2006). Peck (2006) concluded that Business Continuity Management was undertaken primarily for commercial self-interest, rather than to enable continued operations in times of national emergency, and noted that in general preparedness plans for dealing with 'creeping crises' such as pandemic influenza were underdeveloped. Peck (2006) stated "…*for the moment it is unrealistic to assume that Business Continuity Management would ensure the continuity of food and drinks supplies in the event of a national emergency*" and made clear that proactive planning for such an emergency would need to involve government as well as industry, including planning for the suspension of a number of regulations such as the drivers work hours directive. They also note that in considering risk and vulnerability within a supply chain, as opposed to focusing on the interests of a single organisation, there is a need for a wider set of perspectives – integrating more holistic views of risk, embracing macroeconomic visions of the role of the supply chain, and considering complexity arising from an increasingly networked world.

## 2.2. Food supply chain resilience

Some years later, Stone and Rahimfard (2018) argue that exploration of resilience of the agri-food supply chain continues to be a young and fragmented area of research, which would benefit from integration of inputs from multiple disciplines. Through extensive literature review, they map commonly identified elements of resilience to the phases of supply chain resilience identified by Hohenstein et al (2015). They differentiate between resilience of an individual organisation and that of a supply chain, and also between 'core' elements of resilience and 'supporting' elements of resilience (Fig 1).

They further map these elements to the phases of the Adaptive Cycle (Holling & Gunderson, 2001), which is a commonly used as a metaphor in study of complex dynamical systems. This approach contrasts with the traditional engineering definition of resilience, which tends to focus on return to a stable equilibrium, suggesting instead that complex systems go through repeating cycles that can be mapped to development potential, connectedness, and resilience. These cycles can also be nested in a hierarchy, to enable consideration of individual firms and of full supply chain systems. Stone & Rahimifard (2018) propose mapping these against spatial and temporal scales as shown in Fig 2.

Burkhard et al (2011), Fath et al (2015) and Allison & Hobbs (2004) also consider the adaptive cycle in a number of relevant contexts – considering its role in socio-ecological systems and more readily highlighting the role of innovation. Allison & Hobbs (2004) describe traps in the adaptive cycle in the context of Western Australian Agriculture. Rigidity Traps can occur when there is low potential for change, high connectivity and high resilience within a system – it requires a significant external trigger / crisis for a system to tip into a state of radical innovation and subsequent movement to a new growth state. Equally, Poverty Traps can exist in systems that have high potential for change, low resilience and low connectivity – potential for change is not realised due to lack of availability of resources. Fath et al (2015) note that human agency brings an additional complexity not seen in ecological systems, and describe competences that are useful in navigating the adaptive cycle as applied to business management. Burkhard et al (2011) describe how systems can be artificially contained within the Conservation phase through management systems, but that ultimately all phases will end.

Parallels can also be drawn with other cycles integrating business, economics, crisis and innovation, including for example the seminal works of the likes of Schumpeter, Kalecki and Kontradiev. The attraction of exploring the adaptive cycle as an appropriate metaphor for the agri-food system is focused on three significant aspects – (1) it allows for exploration of resilience, and the role of innovation, without explicit connection to the broader economic system; (2) via the concepts of panarchy (Holling et al, 2001), it allows for nesting and hierarchies of cycles; (3) stemming from studies of ecological systems, and later applied to socio-ecological systems, it offers future opportunity to more directly consider the role of the ecological systems and natural resources upon which the agri-food sector is more reliant than most.

## 2.3. Food supply chain and other phases of Resilience and the Adaptive Cycle

Despite the implementation of just-in-time procedures, and other methods to drive operational efficiency, actors in the food supply chain are not strangers to response and recovery from unexpected variation. For actors at the production end of the food chain, small perturbations at the retail end of the chain can have significant impact. Described as the bullwhip effect (Lee et al, 1997), small changes to demand at the retail end of the chain magnify as they propagate through the system, becoming large scale disturbances by the time they reach the producer. For fresh produce these variations can be catastrophic as the long lead time from seed to harvest, combined with short shelf-life of the final product means that growers, already working to tight margins, can be left with excess stock. Similarly, from the supply side, issues such as disease and extreme weather are significant to the fresh produce sector. The bullwhip effect attracts considerable attention from the Operations Management community, and has been observed at sector, organisation and product levels (Wang & Disney, 2016). Wang & Disney (2016) also note that bullwhip can be reduced through supply chain integration, collaboration, information transparency and centralised decisions. Stone & Rahimifard's (2018) literature review and subsequent mapping shows relatively little evidence of research into the elements impacting on the response and recovery phases at an organisational level; however, it highlights collaboration, visibility and agility as core resilience elements at supply chain level. Comparison with the factors supporting reduction of bullwhip suggests that these factors may also support resilience at organisational level. Significant supporting elements are identified to be information flow, cohesion and trust, and communications and bargaining power. Many of these elements have been studied as part of understanding dyadic relationships with the food sector, including within the fresh food sector specifically, and it is commonly noted that there are power imbalances in favour of large scale retailers with producers carrying a greater relative share of risk (e.g. Hingley, 2005). However, determining whether these relationships change in response to crisis appears relatively understudied. Information sharing and collaboration are of particular interest to the Internet of Food Things Network Plus as the enabling potential of digital technology continues to develop (see e.g. Brewer et al, 2019). While power asymmetries no doubt impact on information flow, it is also the case that there are various aspects of competition and anti-trust regulation that inhibit collaboration. Peck (2006) highlighted that response to national emergency may require suspension of some of these regulations, and it is notable that part of the UK Government's response to COVID-19 was to relax competition law – with an associated public policy exclusion order coming into force on 27$^{th}$ March, for a disruption period defined as beginning on 1$^{st}$ March 2020 (UK Gov, 2020). With similar orientation towards increasing collaboration, the Food Chain Emergency Liaison Group first met on March 6$^{th}$, giving opportunity for groups such as Cabinet Office, Public Health England, Food Standards Agency and British Research Consortium (BRC) to share their concerns. It has reportedly met on a weekly basis since. Notably though, the minutes of these meetings are not openly available – leaving most actors within the food supply chain reliant on reports from industry representative bodies and the media.

While our key research questions are focused on resilience of a sub-set of the UK food supply chain, we explore findings in the context of the adaptive cycle, and the role of the Innovation & R&D communities in supporting the supply chain to navigate such a cycle.

## 3. RESEARCH METHODOLOGY

Our empirical research was framed by our three research questions:

1. What was the change in retail demand for fruit and vegetables resulting from COVID-19?
2. What was the impact on UK suppliers of fruit and vegetables?

3. What types of innovation are emerging within supermarket coordinated supply chains of fruit and vegetables?

To address these questions our data was collected in three parts.

1. The data reported are commercial data collected via the Kantar Fast Moving Consumer Goods panel of 30,000 households who report all purchases of food for consumption in the home. We selected products with a diverse range of shelf lives.  COVID-19 first appeared in the UK on 29th January, with the first case of domestic transmission reported on 28th February, and UK Governments 'guide to what you can expect across the UK' published on March 3rd. Closure of the food service sector occurred on 21st March and 'lockdown' was announced on 23rd March.   The data was collected over 12 weeks from week ending 2nd February, and includes weekly volume and value. The data cover the sale of more than £4.8bn of food purchased in the UK over that 12-week period. It should be noted that this does NOT represent product demand since there were many reports of stock outs and rationing throughout the COVID-19 lock down. This simply represents sales (in volume and value).
2. The research team undertook 23 interviews across fruit, vegetable, potato and salad suppliers, specialist organic growers, industry organisations at both local and national level including the National Farmers Union (NFU), the Country Landowners Association (CLA) and the British Growers Association, large scale food processors, food importers, machinery suppliers, glass houses and producers with facilities in the UK and Africa.  The interviews were open and semi-structured covering three broad topics: what changes have you seen in market, what are the impacts on your business and what changes are needed to help you respond? Much of the data was considered highly commercially sensitive as many organisations were still trying to adapt to the COVID-19 changes, and impacts on operational response were live issues.  Therefore, we were unable to record the interviews and instead took contemporaneous notes with follow-up clarifications emailed to respondents where required.
3. Interviews with industry leaders and video workshops with fruit and vegetable suppliers on emerging innovations. These suppliers represented some 50% of the fresh produce community measured by volume.

4. RESEARCH FINDINGS

**4.1. Changes in Retail Demand**

Over the 12-week period commencing 27th January, and 45 lines considered, retail sales saw an uplift of 11% by value and 12% by volume relative to 2019. In the four weeks following shut down of the food service sector, retail sales of the same lines saw an uplift of 14% by value and 13% by volume relative to 2019.  Comparison of variation in value relative to volume shows that pricing strategies varied by product line.  In some cases, prices seem to track volumes on a lagged basis, and in others demonstrate significant volatility that implies a complex mix of demand and supply factors were impacting upon pricing.

Many of the shortage items highlighted in the popular press relate to products with a relatively long shelf life. For items of this type we find an upward lift in sales relative to previous weeks can be seen as early as the week ending March 1st; with uplift relative to 2019 values and volumes often obvious by week ending 8th March.  This is a full 2 weeks prior to the announcement of 'lockdown' and suggests that the UK public were anticipating restrictions in advance of Government guidance. In some cases the uplift in both value and volume, relative to 2019, is significant.

For example, ambient tomato products, such as canned tomatoes, had an uplift by value relative to the same period in 2019 of 103% at the peak of w/e 15$^{th}$ March, and saw week-on-week uplift in sales volume exceed 22% for 3 weeks in a row. They had already begun to see a return to 2019 levels by w/e 22$^{nd}$ (lockdown was announced on 23rd), with week-on-week variation returning to less than 10% by the end of the 12-week period (see Fig 3). It should be noted that 'ambient tomatoes' are cooked and have a shelf life of 18-24 months. There is some evidence of customers making substitution for products that are not readily available – for example, uplift in sales of 'whole ambient tomatoes' occurs 1 week later than the uplift in 'chopped ambient tomatoes', peaking more rapidly with a 55% week-on-week increase in sales by volume.

Similar considerations apply to canned vegetables (113% uplift in value relative to 2019) which peaked a week later than ambient tomatoes, at the w/e 22$^{nd}$ March – still prior to the imposition of lockdown, and following 4 consecutive week-on-week increases in sales volume of 15%-30% (see Fig 4). It is also notable that the trade press began reporting challenges with supply of Italian products, such as ambient tomatoes, at the same time that sales peaked (Grocer, 2020), while the popular press began reporting supermarket rationing of certain products some weeks after sales had peaked (e.g. Guardian, 2020). This highlights that sales figures alone do not necessarily reflect the level of increased demand. Further evidence for more a generalised increase in store cupboard purchasing is provided by the WRAP report (WRAP, 2020) with 59% of shoppers reporting buying more items (+24% for tinned vegetables, +19% frozen vegetables).

Different considerations apply to those in the supply chains of fresh vegetables. Fresh produce is difficult to store and has limited shelf life of 4 to 10 days in retail.

The data for fresh vegetables shows an uplift by value of 35% and by volume of 46% on 2019 for the w/e 22$^{nd}$ Mar. Over the 12-week period, the total volume of fresh vegetables sold was 11% higher than the same period in 2019; albeit this varied across type, with carrots seeing 12% increase relative to 2019 and broccoli having the same demand over 12 weeks in 2020 as it did in 2019.

For standard potatoes, the uplift by value during the peak week was 78% and 54% for new potatoes. It is noticeable that the scale and the timing of demand change for fresh is different from store cupboard items (see Fig 5). The uplift relative to previous weeks does not really begin until the w/e 15$^{th}$ March, peaking the following week (during which lockdown was announced) and falling just below 2019 levels in several categories during the w/e 29$^{th}$ March, rising again towards the end of the period. Interestingly the percentage of consumers reporting waste has fallen from 34% (2018-2019) to 13.7% during the pandemic (WRAP, 2020) suggesting that the closure of food service outlets has resulted in increased consumer food purchasing but with less consumer food waste.

### 4.2. Impact on Food Suppliers

Suppliers reported that the impact of COVID-19 on suppliers of fruit, vegetables, potatoes and salads was very mixed and highly dependent on the specific product. Suppliers in general noted two significant areas that required response: changing demand patterns, and changing availability of labour. Common themes in their ability to effectively respond were focused on flexibility and access to capital assets.

### 4.2.1. Demand Patterns

For suppliers, whose relationships with retailers are largely governed by variable demand call-off contracts, demand patterns relative to 2019 are less significant than week-on-week variation. While the overall supply over a season is constrained by longer term factors such as crop planning and harvest timing, large week-on-week changes cause more operational challenges than those over a whole season as the processes required to pick, pack and deliver are not straightforward to adjust by large amounts. For example, over the 12-week period broccoli saw the same overall demand by volume as 2019, however, the w/e 22$^{nd}$ March saw a 34% week-on-week increase in demand volumes, followed by 21% decrease the following week.

One of our interviewees, who source fresh produce internationally, reported that for the first 18 days of the epidemic, orders were running at 120% and then suddenly "*dropped to zero*".  They noted that their supermarket buyers had left the algorithms that control order replenishment to operate without human intervention, which resulted in supermarkets seeing increased waste as supply exceeded demand. Buyers then began to intervene, manually adjusting the previously automated orders.  This had an enormous knock-on effect for the supplier, who ultimately decided not to trust the order pattern of their customer, to make adjustments based on experience and eventually "*ran our own plan*".  They noted that "*whilst supermarket ordering systems are brilliant in normal phase, they didn't compensate well for large variations emerging because of COVID-19*".  The interviewee concluded that their waste bill was twice its normal level, and that profitability would be reduced accordingly.   This was not an isolated example. An important issue for our supply chain partners was the 'chaos' caused by this algorithmic ordering. Most large retailers use these algorithmic replenishment models and supermarket processes strictly control modifications to subsequent orders, but these huge weekly spikes led to a gross perturbation and accelerated the Forrester Effect (Lee, 1997), with a likely consequence of increased waste throughout the supply chain.

Alongside significant variation in retail demand, sales to the food service sector had collapsed and, in many cases, ceased altogether. As food service is typically 28% of total sales (Defra, 2018) the impact for specialist food service processors has led to closures and widespread furloughing of staff in food service and retailing. The impact on the potato sector (and a few other products such as lettuces) is significant as many suppliers have big markets linked to fast food and catering – for example, in 2019 the service sector accounted for 36% of the domestic potato processing market (AHDB, 2020a).  Potato processors who were equipped to shift their operations towards the retail market were able to cope with changing demand, where those with specialised operations and associated cost bases were faced with volume collapse.  For example, the closure of fish and chip shops leaves the industry estimating that this has left 200,000 tonnes of product without a market (AHDB, 2020b).  The result is that UK prices have fallen and suppliers are reducing UK crop planting by circa 20% (AHDB, 2020b).

**4.2.2. Labour Impacts**

The major concern for our interviewees, both farmers and growers/packers, is that the rules brought in to contain COVID-19 have reduced labour productivity by "*typically 30%*" as a consequence of social distancing in factories and the requirement to transport workers.  Other costs have risen including recruitment, training, acquiring Personal Protective Equipment and cleaning. Suppliers reported that costs have risen, but they have not been able to pass these costs on, meaning in many cases that increased turnover is leading to reduced profits or even to trading losses.  The other major strategic issue for vegetable and salad companies is a concern that the ways in which they have met short term demands for extra staff, including using furloughed staff, increased overtime etc., cannot be sustained; and there are concerns about access to migrant labour over the longer term.

COVID-19 produced a scale of change on labour supply far beyond what is normal. Across our whole set of companies those who supplied the food retail market, saw retail demand increase typically by 10-15% across the 12-week period, whilst those who normally supply food outlets were operating at less than 10% of normal capacity because of the closure of restaurants, pubs and fast food outlets. At the extreme end one of our companies had a six-fold increase in demand in a four-week period. The potato farmers who supplied restaurants saw an almost complete loss of demand in two weeks whilst farmers of new potatoes saw retail demand rise by 48%. For those adapting to increased demand, at the same time as they were experiencing rapid rises they reported increasing staff absentee rates as staff self-isolated, either to protect themselves or due to illness or susceptibility in their household and/or family. This exacerbated the problem of meeting large changes to demand.

### 4.3. Emerging Innovations

#### 4.3.1. Demand Patterns

In some instances there was evidence that suppliers were able to adapt rapidly to changing market conditions – these were predominantly situations in which suppliers were already focused on the retail sector, and were primarily dealing with uplift in demand. Suppliers usually dealing with the service sector coped better in situations in which there were not significant capital assets (such as packaging equipment) required in order to enable them to switch product line (e.g. to create smaller volume packages to deliver to retail rather than wholesale) or channel (e.g. responding to increased demand from online retailers, or delivering direct to consumers via 'veg boxes'). In situations where this was not possible the role of convening groups, representative bodies and levy boards has been visibly important. For example in early May, building on their open-access analysis of potato markets, AHDB launched 'The Potato Portal' as a mechanism by which to connect potato growers with wholesale buyers (AHDB, 2020c).

#### 4.3.2. Labour – immediate response

In the short term this problem was being addressed through the Pick for Britain platform. This was developed as a collaborative venture between AHDB, the Association of Labour providers, British Growers, Department for Environment Food and Rural Affairs, the National Farmers Union (NFU) and Greater Lincolnshire Local Enterprise Partnership. One of our interviewees had been heavily involved in developing the campaign and saw it as an example of "*extraordinary industry collaboration*" driven by "*an emerging crisis*" in labour. At the outset of the pandemic he reported that each organisation was developing their own approach to the problem, but with the National Farmers Union taking the lead each of the main organisations brought these elements together to specify a single platform.

Government intervention also played an important role in addressing labour challenges. Following pressure from a number of sectors experiencing labour challenges, Government amendment to the Coronavirus Job Retention Scheme allowed furloughed workers to work for another employer whilst still being furloughed. In some cases, this made a significant difference in fulfilling the demand for pickers, factory workers and drivers. One of our respondents reported resorting to social media to advertise their labour requirements and quickly picked up furloughed staff from other industries. More typically, firms relied on existing agencies.

Further Government intervention took place in the form of temporary relaxation of the enforcement of drivers' hours rules. Starting on 18$^{th}$ March and ending on 16$^{th}$ April, this relaxation applied solely in the supply of food and other essential products to supermarkets, including the movement of such goods from importers, manufacturers and suppliers to distribution centres. It did not apply to drivers undertaking deliveries directly to consumers, but provided some relief to a number of suppliers.

#### 4.3.3. Labour – longer term adaptation

In the video workshops we focused on addressing the longer-term developments for overcoming the labour shortages. The Internet of Food Things Network Plus facilitated these four workshops, with over 80 companies and organisations who represented over 50% of the fresh produce community by volume. There was wide scale recognition across the industry that there is a growing need to accelerate its approach to reducing dependencies on seasonal migrant workers.  The industry focused on opportunities for robotics to drive labour productivity in the short and long term, reduce work to worker COVID-19 transmission risk, and upskill the existing permanent labour force.  Robotics and automation companies who participated in the workshops noted the significant export potential for these technologies, and opportunity for the UK to lead in demonstrating the efficacy of these solutions. There was considerable interest across four main areas: apple and strawberry harvesting service robotics; blueberry picking; broccoli harvesting; and packing systems using industrial robots. This included the development of a fast-track process for patents to secure the collaboration. Both the broccoli harvesting and the packing systems required industrial robot systems and the development of the specification was in collaboration with the Manufacturing Technology Centre, which was established in 2010 as an independent research and technology organisation with the explicit aim of bringing together academia and industry. These workshops resulted in a large consortium bidding to DEFRA and UKRI for funding, and provision of evidence to a Parliamentary Select Committee seeking suggestions for policy support to mitigate COVID-19 impacts on food security.

## 5. IMPLICATIONS & FUTURE WORK

This case study based exploratory research, considering the impact of COVID-19 on the supply chain for fruit and vegetables within the UK, has implications for management theory and practice.

The empirical evidence that we have collected can be applied to the theoretical framework developed by Stone & Rahimifard (2018), and to an adapted framework based on their work and that of Burkhard et al (2011), Fath et al (2015) and Alison & Hobbs (2004).  Figure 6 shows a framework in which the adaptive cycle is mapped to the resilience phases identified by Stone & Rahimifard (2018), but which also contains a nested cycle in the event that the trigger or shock to the system is not sufficient to tip the system into a new state of operation.  This also incorporates the concept of a rigidity trap in which there is an inability to exploit new innovations to enable new modes of operation.

The UK food system is often characterised as highly complex with many each tier of the supply chain locked into current working model that delivers food cheaply to the consumer. Despite a long history of calls for change around, for example, emerging markets such as local food initiatives (e.g. Brewer et al, 2019) the potential for development is limited by the procurement practices of the large UK supermarkets who together account for £157Bn of the £179Bn grocery retail market value (Kantar, 2017). The system also exhibits high connectivity with farmers, producers and logistics tied into contracts and call offs that provide outlets for their production. In the context of Figure 6, it is likely that pre-COVID, the UK food system sat in the upper right 'Status Quo' section – highly interconnected with limited potential for change. The key question is how resilient is the system?

Following a very clear 'trigger' (in some cases demand rose by over 100% on a year-by-year basis and over 50% in two weeks) we saw evidence that the core-elements identified by Stone & Rahimifard (2018) for the Response Phase at supply-chain level resilience are also relevant at firm and tier levels within the UK fruit and vegetable supply chain.  In addition, we note that for system shocks that exist and evolve over a time period of several weeks, or longer, there is no clear delineation between the phases. In particular, 'Readiness' and 'Response' phases have significant overlap, and many of our interviews demonstrated that suppliers were undertaking planning activities alongside operational response.  These actions of 'learning' are traditionally undertaken within the 'Adaptation' phase, and suggest that in considering 'Readiness' managers should be aware that phases are likely to run in parallel. In diagrammatic form we consider these as cycles nested within a broader cycle, as illustrated in Figure 6.

During the Response phase we saw that many of the issues highlighted by Peck (2006) in their rather prescient report for DEFRA, came to fruition – including labour shortages and high levels of absenteeism, 'panic buying', sharp rises in requirements for home delivery and the potential for reduced variety amongst manufacturers. This represents a clear and direct contribution from the research and innovation community, notably:

- Peck highlighted that the Drivers working hours directives may need to be relaxed – this occurred between March 18$^{th}$ and April 16th
- Peck highlighted that benefit rules may need to be amended to allow staff to work longer than usual hours without being penalised by loss of benefit entitlement – whilst this has not been implemented directly, there are parallels with the amendments to the Coronavirus Job Retention Scheme.
- Peck highlighted that anti-trust regulations that inhibit competitors from collaborating may need to be relaxed. This has been implemented, and while our interviews did not provide evidence that this has contributed to increased sharing of information or collaborative activity, it clearly represents a potential area for future exploration when considered alongside the demand driven impacts on algorithmic ordering systems.

Some aspects that they highlighted as possible responses have not yet been seen – such as the requirement for policing in stores – but our findings align closely with their report. We note that despite a relatively significant trigger for disruption, the organisational response was strongly limited to modifying existing procedures and practices. There were some limited examples of innovative practice including the coming together of many industry bodies in the development of the Pick for Britain platform and the Potatoes Portal. There was also a willingness to engage with research institutions such as the IoFT Network Plus who acted as a broker for consideration of more radical innovations around the use of robotics.

We believe that in this paper we have provided evidence that despite newspaper headlines about food shortages, the COVID-19 shock has shown a system that, from an operational perspective at least, is resilient to both demand and supply side (e.g. labour) problems. This combination points to a system that is stuck in the 'rigidity trap'. It appears that despite the severity and suddenness of the onset of COVID-19, it was still not sufficient of a shock to move the industry out of its existing structure. There is some limited evidence of the industry considering moving out of the 'Status Quo' and imagining an alternative future that addresses longer term systemic challenges, but these developments seem very long-term and reliant on funding from actors outside of the operational supply chain such as government research and innovation agencies.

This work highlights the potential for a future inter-disciplinary research agenda of significant value to society, national economies, and individual firms. Parallels between the food supply chain, the adaptive cycle and panarchy warrant further exploration – there are lessons to be learned from the socio-ecological systems community, and the innovation community has a role to play in supporting imagining of alternative states for both the food supply chain as a whole and individual tiers and firms within it. Possible contributions from the systems dynamics community include the quantification of system thresholds and tipping points, enabling the supply chain management community to undertake more effective preparation and planning for specific phases in the cycle. Similarly, increasing automation of ordering processes will require data scientists to explore how significant perturbations can be integrated or over-ridden. Bringing together these varying perspectives, ensuring common language and understanding, and enabling effective participation of communities with differing goals and objectives, is no small task.  There is a significant role to be played by the innovation and R&D communities, including public & private R&D labs and communities of practice, in facilitating and driving these activities.

## 6. CONCLUSION

COVID-19 has been a significant shock to the food supply chain. In the UK this led to the closure of food service outlets and at the same time a significant increase in retail purchasing. For suppliers of fresh produce uplifts of the scale seen are very challenging, particularly when supply was predicated on minor adjustments to 2019 or week-on-week volumes. However, it wasn't just a one-off increase. Because of the closure of food service outlets some products have seen a sustained increase whilst others saw the same overall demand by volume as 2019, and significant variation on a week-by-week basis.

Following our interviews with 23 organisations, we have identified some generalised findings that relate to the special issue focus on finding solutions in emergencies. Firstly, a number of new organisational procedures emerged including adaptations to ordering algorithms to reduce the Bullwhip effect and suppliers developing their own production schedules independent of retail orders. Secondly, new ways of collaborating emerged including the Pick for Britain and Potato Portal emerged in very short time frames involving multiple semi-public bodies, government agencies and departments and firms working together to develop solutions. Innovation efforts also included working with government to influence policy for example on the Coronavirus Job Retention Scheme which enabled those producers seeing rapid increases in demand able to use surplus labour form the food service sector. In the longer term there are indicators that the sector is coming together to work on the issue of long-term labour shortages. The Internet of Food Things Network Plus ran of series of workshops to identify how robotics could be used to relieve the labour shortage and these innovations are now under active consideration by public funders.

Finally, COVID-19 led many commentators to promote the idea of a longer-term structural change in the industry in support of local production and markets. Our findings indicate a short-term shock but that this shock is insufficient to move the industry out of its rigidity trap. As strong as the COVID-19 shock was, the sector's resilience, despite very little redundancy, appears to have kept the industry firmly in its status quo. Radical innovations such as industrial robotics have been energised but it may require a second, or prolonged, emergency to further these developments.

**FIGURES**

**Fig 1.** *Tabular version of Stone & Rahimifard (2018) mapping of Context, Phase and Core Resilience Elements of supply chain resilience strategies.*

| Context | Phase | Resilience Core Elements |
|---|---|---|
| Organisational Resilience | Readiness Strategy | Early Warning Systems<br>Redundancy<br>Flexibility<br>Security |
| | Response Strategy | |
| | Recovery Strategy | |
| | Adaptive Strategy | Risk Aware Culture |
| Supply Chain Resilience | Readiness Strategy | Redundancy<br>Flexibility |
| | Response Strategy | Collaboration<br>Visibility |
| | Recovery Strategy | Agility |
| | Adaptive Strategy | Adaptability |

**Figure 2:** *Reproduced from Stone & Rahimifard (2018) – the agri-food supply chain represented as an adaptive cycle, alongside representation of phases of resilience.*

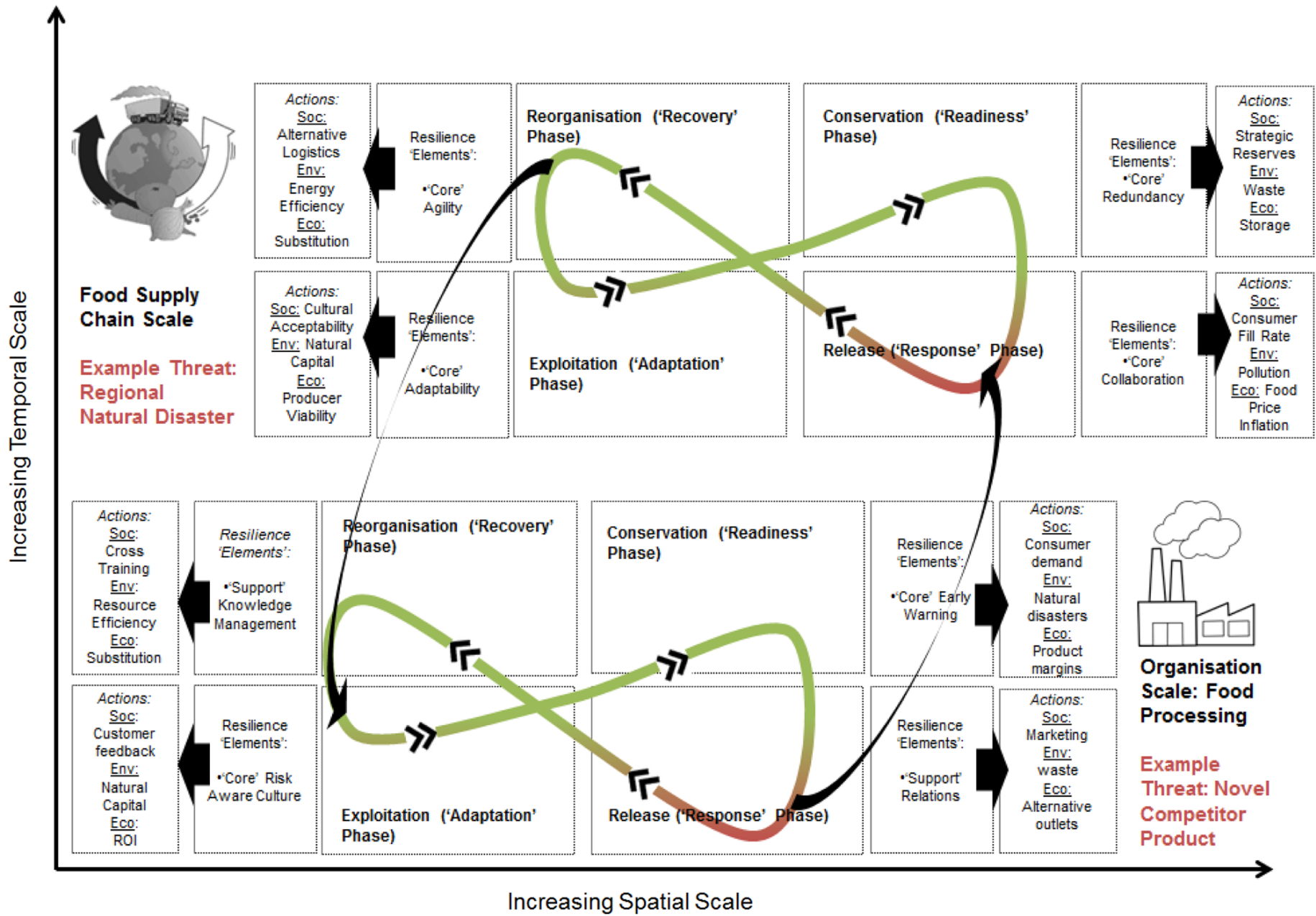

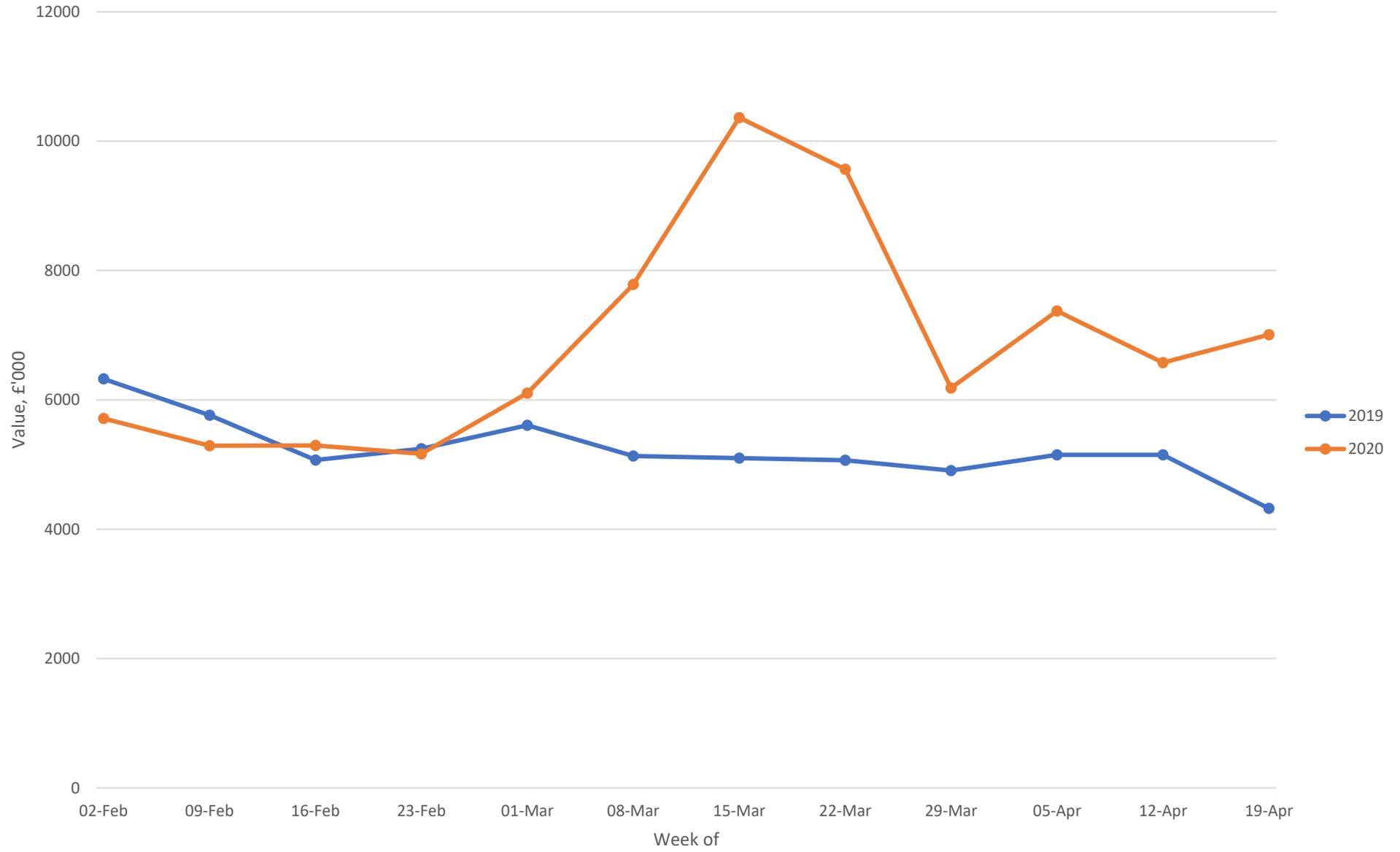

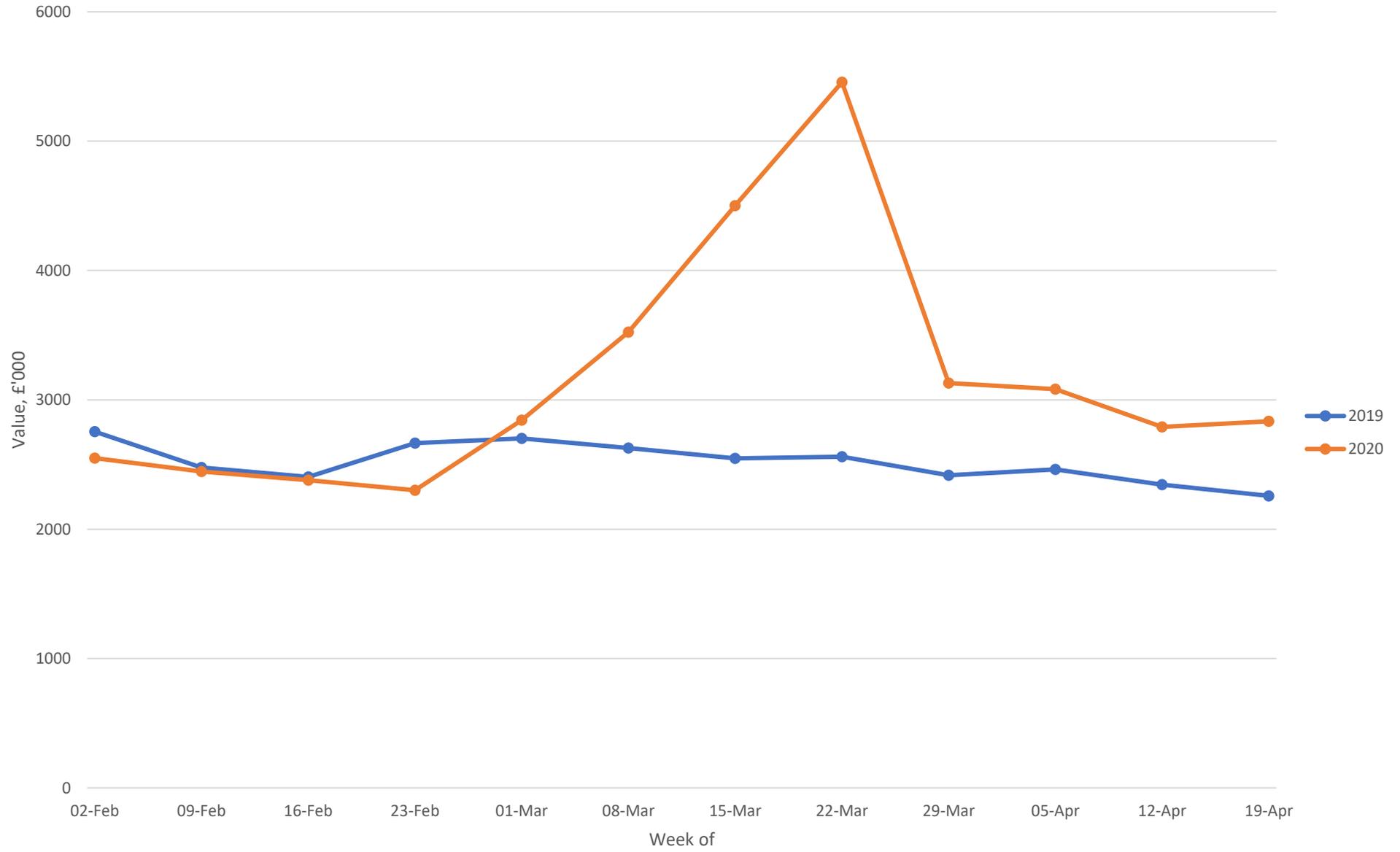

Figure 4: Sales of Canned Vegetables

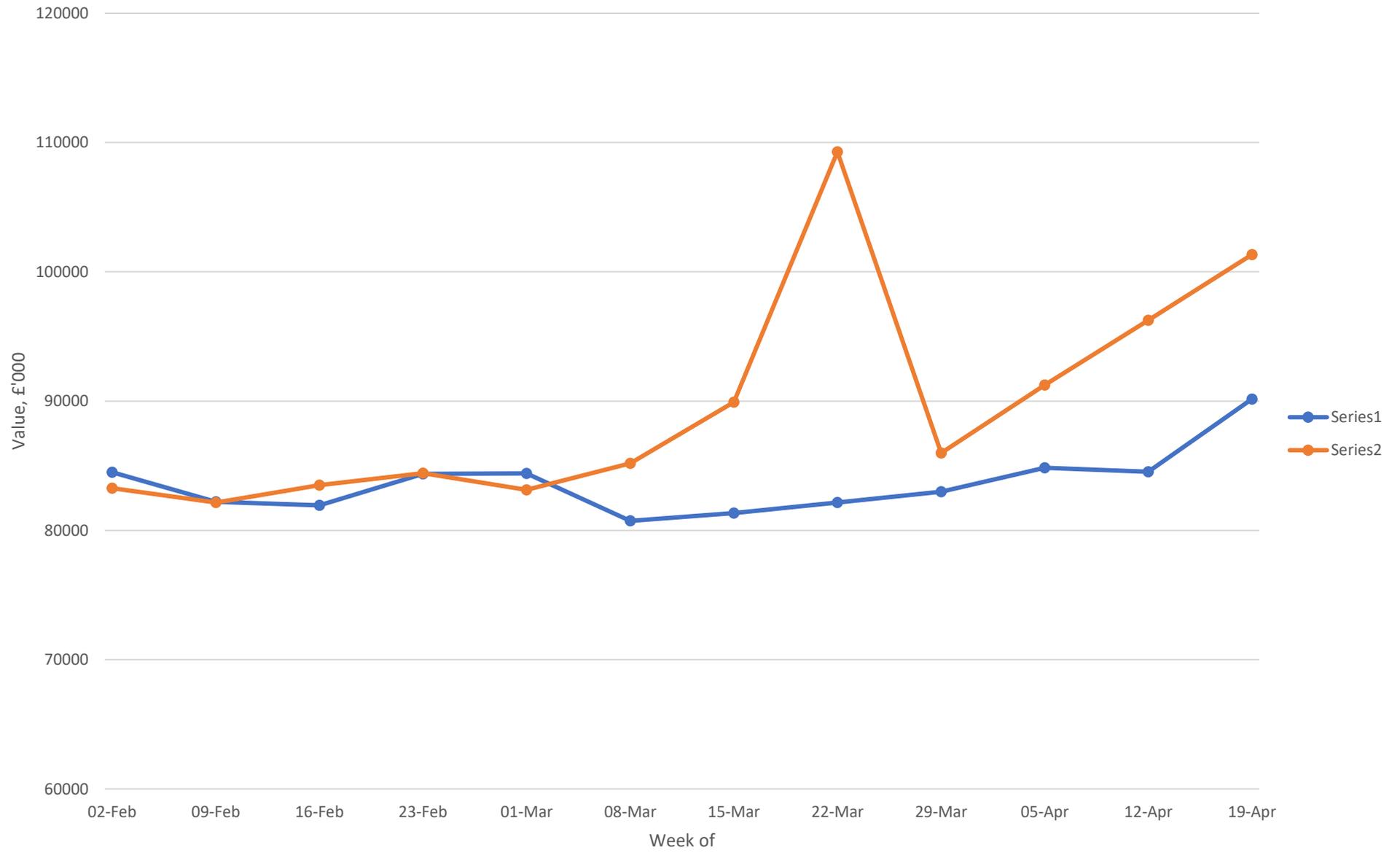

Figure 5: Sales of Fresh Vegetables - inc broccoli, carrots, lettuce

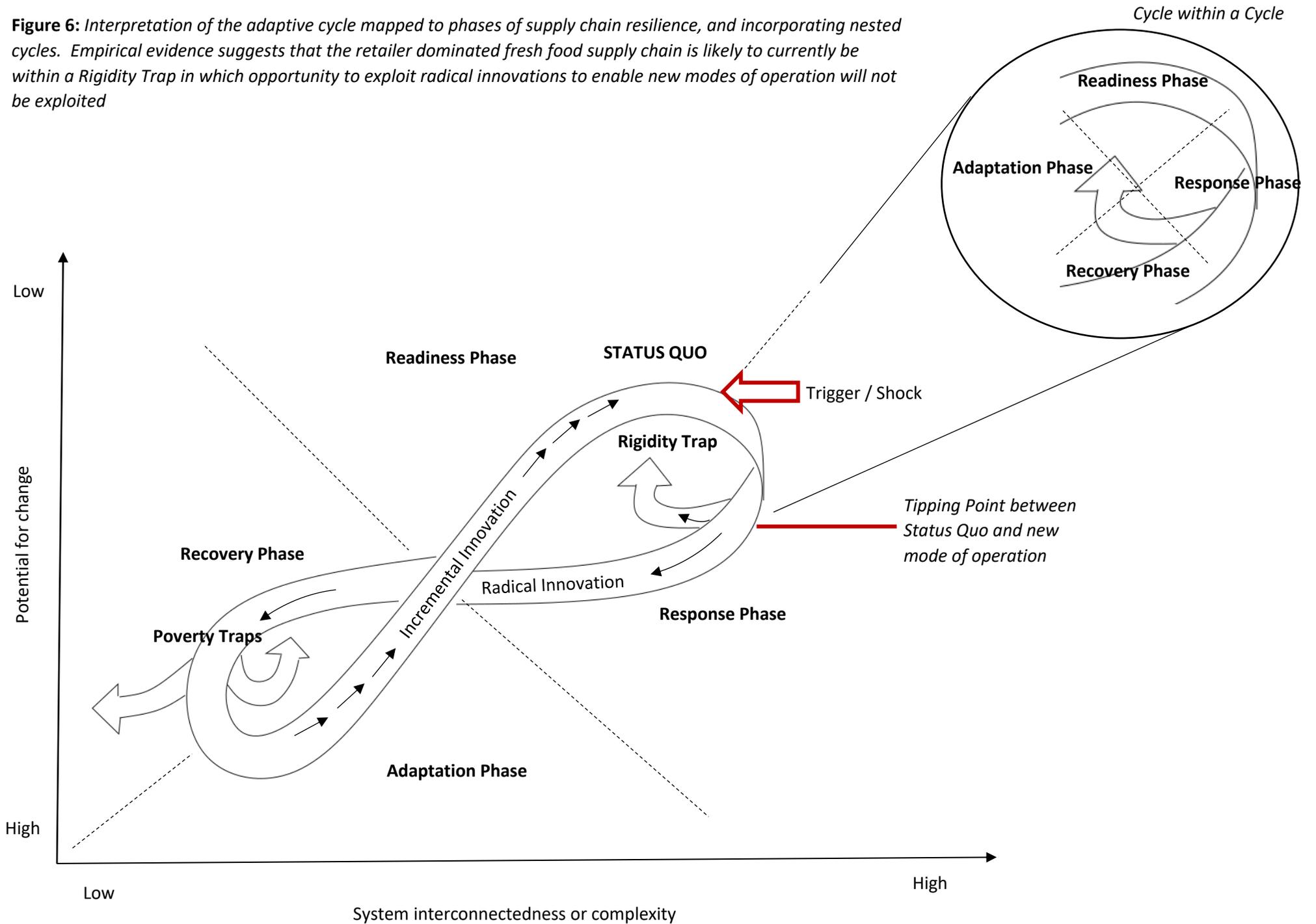

**Figure 6:** *Interpretation of the adaptive cycle mapped to phases of supply chain resilience, and incorporating nested cycles. Empirical evidence suggests that the retailer dominated fresh food supply chain is likely to currently be within a Rigidity Trap in which opportunity to exploit radical innovations to enable new modes of operation will not be exploited*